\documentclass[eqsecnum,aps,twocolumn]{revtex4}
\usepackage{graphics}

\begin{document}
\title{\sc On Mathematical Defect in Demonstration of Convection Theorem:
Implications For Continuum Mechanics and Other Classical Field
Theories}

\author{R. Smirnov-Rueda}
\address{Applied Mathematics Department, Faculty of Mathematics,
Complutense University, 28040, Madrid, Spain}

\date{\today}
\begin{abstract}
Serious mathematical defect in the important kinematics theorem
known in continuum mechanics as Convection (or Transport) Theorem
is reported. We claim that the traditional demonstration does not
take into account a special constraint on integrand functions
given in Lagrangian representation. Thus, we put in doubt that the
traditional procedure for the transition from integral
formulations of physical laws of some classical field theories to
their differential form is mathematically rigorous. Reconsidered
formulation shows the way how the system of fundamental
differential equations of continuum mechanics and some other field
theories could be modified without any change of the original
integral formulation. The continuity equation and the differential
form of the equation of motion for continuous media are discussed
as examples of modification.
\end{abstract}
\pacs{} \maketitle


\section{Introduction}

At present, a huge amount of experimental data has been accumulated on
deformable bodies, fluids, gases and plasmas. Mathematical understanding of
their behavior, internal relationship between different concepts, models and
observed phenomena are expressed in comprehensive modern descriptions
generally referred as hydrodynamics, elasticity theory, electromagnetism,
magnetohydrodynamics, plasma physics (as regards to the latter, nowadays a
special priority, attention and technical support have been given to the
solution of problems of thermonuclear reaction as alternative source of
energy). The most part of inner links between these autonomous branches of
scientific knowledge is mainly due to the common mathematical apparatus of
continuum mechanics which brings them together as a part of a more general
scheme.

Having in mind this unity of mathematical basis for all
above-mentioned subdivisions of physics and mechanics, we shall
refer in this work only to a mechanical theory of motion of
continuous media which constitutes a significant part of continuum
mechanics. The development of the theory of ordinary and partial
differential equations, integral equation, differential geometry
etc had great influence on conceptual, logical and mathematical
structure of continuum mechanics and \textit{vise versa}. In
historical retrospective, it is now commonly accepted that the
consolidation of mathematical and conceptual fundamentals of the
theory of continuum mechanics was achieved by the end of the 19th
century and, in main terms, coincided with the rigorization of the
analysis completed by Weierstrass [$1$].


However, very recent indications [$2$] on ill-founded analysis in
mathematical hydrodynamics put in doubt the fact that the process
of rigorization of the fundamentals of classical field theory had
been brought to the end in a correct way. More precisely, a
detailed insight towards Lagrangian and Eulerian types of
analytical description conventionally accepted for kinematics of
continuous media shows that no equal standards of rigor are
implied in both approaches to time derivatives. A reconsidered
account [$2$] provided a mathematically rigorous analytical
approach to the treatment of total time derivative in properly
Eulerian description.\ Another serious defect had been detected in
the demonstration of an important kinematics theorem known also as
Convection (or Reynolds' Transport) Theorem [$2$]. Elimination of
both defects provided a necessary cross-verification for a
modified differential form of continuity equation [$2$].

In what follows, one of our immediate purposes will be to fix basic
conceptions of mathematical foundations of continuum mechanics and then we
shall proceed to show the failure in the logic of the traditional approach
to the formulation of the Convection Theorem and how its reconsideration
motivates a new form of differential equations of motion for continuous
media.


\section{Basic Descriptions in Continuum Mechanics}

The basic mathematical idea of kinematics is that a continuous
medium can be consistently conceived as an abstract geometrical
object (for instance, collection of spatially distributed points
etc). Thus, any deformation and motion is immediately associated
with appropriate geometrical transformation [$3$]-[$4$]. Since any
motion is always determined with respect to some reference frame,
let us introduce a fixed coordinate system. To simplify our
approach in order to highlight the nature of the defects inherent
in the conventional demonstration of the Convection Theorem, we
begin with the simplest assumptions of continuous geometrical
transformations.

Actual mathematical formalism implies two complementary general descriptions
of flow field kinematics. One of them, called \textit{Eulerian description
(or representation}), identifies a fixed volume element or a fixed point of
space in a chosen reference system. All medium properties are described only
as a function of local position $\mathbf{r}$ and time $t$. These independent
variables are frequently regarded as \textit{Eulerian variables}.

The other approach, called \textit{Lagrangian description}, identifies an
individual bit of continuous medium (characterized by an initial closure $%
\Omega _{0}$) or a point-particle (characterized by an initial
position-vector $\mathbf{a}$) at some chosen time instant and gives account
of the medium properties along their trajectory. This approach associates
non-zero motion with a non-zero continuous geometrical transformation $H_{t}$%
. Thus, the set $H_{t}\Omega _{0}$ (or a position-vector $H_{t}\mathbf{a}$)\
represents the same individual bit of continuous medium (or point-particle)
at time $t$. Continuity requirement on $H_{t}$ involves a natural
limitation: the bounding surface of the closure $H_{t}\Omega _{0}$ always
consists of the same medium elements, i.e. $\partial \Omega
_{t}=H_{t}\partial \Omega _{0}$ and there is no flux of medium particles
through the boundary at any instants of time. Thus, under this conditions
the transformation $H_{t}\mathbf{a}$ is always the \textit{point
transformation }and\ the function\textit{\ }$H_{t}\mathbf{a=r(}t,\mathbf{a})$
describes a law of motion of a point. The position-vector $\mathbf{a}$
denotes the initial position of an individual particle and, therefore, can
be used as \textit{a label} for constant identification of the particle at
any instant of time. An initial set of identified particles is equivalent to
the set of labels $\left\{ \mathbf{a}\right\} $ regarded sometimes as
\textit{Lagrangian parameters}. The assumption on continuity of geometrical
transformation $H_{t}$ is equivalent to the requirement that the function $%
\mathbf{r(}t,\mathbf{a})$, which describes the law of motion, possesses
continuous partial derivatives with respect to all variables, i.e. $t$ and $%
\mathbf{a}$.

In considering the motion of continuous medium as a set of
individual mutually interacting point-particles (or volume bits),
Lagrangian approach is indispensable as the first step, implying
by it the individualization of particles by a set of labels
$\left\{ \mathbf{a}\right\} $. Thus, it should be emphasized again
that the geometrical transformation $H_{t}\mathbf{a=r(a,} t)$
gives the complete picture of a motional history for every
individual particle from the set $\left\{ \mathbf{a}\right\} $.
The detailed description of the law of motion implies an
introduction of certain additional concepts such as the velocity
and acceleration of particles of a continuous medium. In
Lagrangian description velocity and acceleration are defined as
the first and the second order partial time derivative with
respect to $\mathbf{r}$, respectively:

\begin{equation}
\mathbf{v=}\frac{\partial }{\partial t}\mathbf{r(}t,\mathbf{a});\qquad \frac{%
\partial \mathbf{v}}{\partial t}=\frac{\partial ^{2}}{\partial t^{2}}\mathbf{%
r(}t,\mathbf{a})  \label{odin}
\end{equation}

In the context of Eulerian description, \textit{a priori} there is no
identification and hence no explicit consideration of the function $\mathbf{r%
}=\mathbf{r(}t,\mathbf{a})$. The primary notion is the velocity field as a
function of position $\mathbf{r}$ in space and time $t$ on some domain of a
continuous medium:

\begin{equation}
\frac{d\mathbf{r}}{dt}=\mathbf{v(}t,\mathbf{r})  \label{dva}
\end{equation}%
where variables $\mathbf{r}$ and $t$ are independent.

Picking up some initial point $\mathbf{a}=\mathbf{r(}t_{0})$, one selects
from a congruence (a set of integral curves of (\ref{dva})) a unique
solution. Thus, a formulation of the \textit{initial Cauchy problem}

\begin{equation}
\frac{d\mathbf{r}}{dt}=\mathbf{v(}t,\mathbf{r});\qquad \mathbf{r(}t_{0})=%
\mathbf{a}  \label{tri}
\end{equation}%
is mathematically equivalent to an act of identification, allowing any
solution of (\ref{tri}) to be represented as in Lagrangian description $%
\mathbf{r=r(}t,\mathbf{a})$. There is a general consensus that
this procedure can be taken as a rule for translating from one to
the other description. Thus, if some medium quantity $f$ is
defined in Eulerian representation as $f(t,\mathbf{r})$, then
there is an obvious translation rule to its Lagrangian
representation [$4$]:

\begin{equation}
g(t,\mathbf{a})=f(t,\mathbf{r(}t,\mathbf{a}))  \label{chet}
\end{equation}

\textit{Convective (or Euler's material) derivative} is introduced as:

\begin{equation}
\frac{\partial }{\partial t}g(t,\mathbf{a})=\frac{d}{dt}f(t,\mathbf{r(}t,%
\mathbf{a}))=\frac{Df}{Dt}=\frac{\partial f}{\partial t}+(\mathbf{v\cdot
\nabla })f  \label{pyat}
\end{equation}

Now we are in a position to consider an important kinematics
theorem which concerns the time rate of change of $f$-content of
any volume integral (i.e. not only infinitesimal volume elements).
Its formulation can be found in any basic text on fluid dynamics
or elasticity theory. For our convenience, we shall use the
exposition and symbolic notations implemented in [$4$]. As the
first step, let us define $f$-content of some deformable moving
volume domain $\Omega _{t}$ and its time derivative in Eulerian
and Lagrangian representations, respectively:

\begin{equation}
F(t)=\int\limits_{\Omega _{t}}f(t,\mathbf{r})dV;\qquad \frac{dF}{dt}=\frac{D%
}{Dt}\int\limits_{\Omega _{t}}f(t,\mathbf{r})dV  \label{shest1}
\end{equation}
and

\begin{eqnarray}
F(t) &=&\int\limits_{\Omega _{t}}f(t,\mathbf{r(}t,\mathbf{a}%
))dV=\int\limits_{\Omega _{t}}g(t,\mathbf{a})dV;   \nonumber \\
\frac{dF}{dt} &=&\frac{\partial }{\partial t}\int\limits_{\Omega _{t}}g(t,%
\mathbf{a})dV  \label{shest2}
\end{eqnarray}
where $\Omega _{t}=H_{t}\Omega _{0}$.

The geometrical transformation $H_{t}$ is algebraically represented by the
Jacobian determinant $J=\det \left\vert \frac{\partial H_{t}\mathbf{a}}{%
\partial \mathbf{a}}\right\vert $ which has the following partial time
derivative [$4$]:

\begin{equation}
\frac{\partial }{\partial t}J=\frac{\partial }{\partial t}\det \left\vert
\frac{\partial H_{t}\mathbf{a}}{\partial \mathbf{a}}\right\vert =(\mathbf{%
\nabla \cdot v})J  \label{sem}
\end{equation}%
where $(\mathbf{\nabla \cdot v})=div\mathbf{v}$.

Thus, in the framework of Lagrangian representation the evolution of a
medium $f$-content can be written in original variables $\mathbf{a}$, i.e.
the integration is taken over the initial volume domain $\Omega _{0}$:

\begin{equation}
\frac{dF}{dt}=\frac{\partial }{\partial t}\int\limits_{\Omega _{0}}g(t,%
\mathbf{a})JdV_{0}=\int\limits_{\Omega _{0}}\frac{\partial }{\partial t}(g(t,%
\mathbf{a})J)dV_{0}  \label{vosem}
\end{equation}%
where $g(t,\mathbf{a})$ should be evaluated in the volume set
$\Omega _{0}$, i.e. under the condition
$\mathbf{r}(t,\mathbf{a})=\mathbf{a}$. In fact, this is only
meaningful just for the initial configuration of point-particles
$\left\{ \mathbf{r}(t_{0},\mathbf{a})=\mathbf{a}\right\}$ which
continuously fill up the original domain $\Omega _{0}$. Therefore,
taking into account this constraint

\begin{equation}
\left[ g(t,\mathbf{a})\right] _{\mathbf{r}(t,\mathbf{a})=\mathbf{a}}=\left[
f(t,\mathbf{r(}t,\mathbf{a}))\right] _{\mathbf{r}(t,\mathbf{a})=\mathbf{a}%
}=f(t,\mathbf{a})  \label{vosema}
\end{equation}%
we can proceed to evaluate the following partial time derivative:

\begin{equation}
\frac{\partial }{\partial t}(g(t,\mathbf{a})J)=\frac{\partial }{\partial t}%
(f(t,\mathbf{a})J)=\frac{\partial f}{\partial t}J+f(\mathbf{\nabla \cdot v})J
\label{devyat}
\end{equation}%
and it suffice to note that

\begin{equation}
\frac{dF}{dt}=\int\limits_{\Omega _{0}}\left[ \frac{\partial f}{\partial t}%
+f(\mathbf{\nabla \cdot v})\right] JdV_{0}  \label{desyat}
\end{equation}%
which lead to the modified formulation of the Convection Theorem:
\newtheorem{LCT}{Theorem}
\begin{LCT}{:} Let $\mathbf{v}$ be a vector field
generating a fluid flow through a fixed 3-dimensional domain $V$ and
if $f(\mathbf{r},t)\in C^{1}(\bar{V})$, then

\begin{equation}
\frac{D}{Dt}\int\limits_{\Omega _{t}}fdV=\int%
\limits_{\Omega _{t}}(\frac{\partial f}{\partial t}+f(\mathbf{\nabla
\cdot v}))dV \label{odinadzat}
\end{equation}
where $dV$ denotes the fixed volume element.

\end{LCT}

The formulation of (\ref{odinadzat}) differs form the
conventionally accepted result. The defect in the traditional
demonstration resides in the fact that the relation (\ref{vosema})
was not taken into account (for instance, see [$4$]). In fact, if
we take a partial time derivative (\ref{devyat})
straightforwardly, i.e. without the constrain (\ref{vosema}), we
obtain:

\begin{equation}
\frac{\partial }{\partial t}(g(t,\mathbf{a})J)=\frac{\partial }{\partial t}%
(g(t,\mathbf{a})J)=\frac{Df}{Dt}J+f(\mathbf{\nabla \cdot v})J
\label{devyata}
\end{equation}
that leads to the traditional formulation of the Convection
Theorem [$4$]:
\begin{equation}
\frac{D}{Dt}\int\limits_{\Omega _{t}}fdV=\int\limits_{\Omega _{t}}(\frac{Df}{%
Dt}+f(\mathbf{\nabla \cdot v}))dV  \label{odinadzata}
\end{equation}%
where we remind that the right-hand side of the expression
(\ref{odinadzat}) or (\ref{odinadzata}) implies the condition
$t\rightarrow t_{0}$, i.e. $\Omega _{t}\rightarrow \Omega _{0}$
and corresponding integrands are represented in Eulerian variables
attached to the fixed spatial reference system. Thus, to end this
Section, we conclude that the Convection Theorem as well as the
demonstrational procedure it involves are important mathematical
tools in order to describe relationship between Lagrangian and
Eulerian representations for the set of fundamental equations of
continuum mechanics.


\section{Conservation Law and Modified Differential Equation of Motion}

One of the primary fundamental equations of continuum mechanics is the
differential equation for the law of conservation of the mass of any closed
volume element, i.e. \textit{continuity equation}. Let us choose the
modified version of the Convection Theorem:

\begin{equation}
\frac{D}{Dt}\int\limits_{V_{t}}\rho (t,\mathbf{r}%
)dV=\int\limits_{V_{t}}\left[ \frac{\partial \rho }{\partial t}+\rho \,div(%
\mathbf{v)}\right] dV=0  \label{dvenadzat}
\end{equation}

Hence, the first modified fundamental differential equation of
continuum mechanics given in Eulerian representation would take
the form

\begin{equation}
\frac{\partial \rho }{\partial t}+\rho \,div(\mathbf{v)}=0  \label{trinadzat}
\end{equation}%
where $\rho $ is the mass density. The equation (\ref{trinadzat})
can be regarded as a \textit{modified continuity equation} which
was already independently obtained due to reconsidered approach to
total time derivatives in properly Eulerian description [$2$].
Mathematical soundness and applicability of the equation
(\ref{trinadzat}) was also analytically verified on a simple
one-dimension example of an ideal flow (see Appendix B in [$2$]).

In continuum mechanics, there are many physical quantities such as
linear and angular momentums, energy as well as some other scalar,
vector or tensor characteristics which undergo time variations
during the motion of any given volume of a medium. If this
quantities are continuous functions of the coordinates everywhere
inside the spatial domain $V_{t}$, then the mathematical procedure
used for the demonstration of the Convection Theorem
(\ref{odinadzat}) also remains valid. Thus, for any moving
individual macroscopic volume $V_{t}$, the equation for time
variation of linear momentum will take the following form for each
spatial component $v^{i}$ of the velocity vector $\mathbf{v}$:

\begin{equation}
\frac{D}{Dt}\int\limits_{V_{t}}\rho v^{i}dV=\int\limits_{V_{t}}\left[ \frac{%
\partial \rho v^{i}}{\partial t}+\rho v^{i}\,div(\mathbf{v)}\right] dV
\label{chetyrn}
\end{equation}

The concept of force is introduced in continuum mechanics phenomenologically
by analogy with classical mechanics. In other words, different forces which
act on the volume $V_{t}$ are responsible for the time variation of momentum:

\begin{eqnarray}
\frac{D}{Dt}\int\limits_{V_{t}}\rho \mathbf{v}dV=
\int\limits_{V_{t}}\left[ \frac{\partial \rho \mathbf{v}}{\partial
t}+\rho \mathbf{v}\,div(\mathbf{v)} \right]
dV=\nonumber \\
\int\limits_{V_{t}}\rho \mathbf{f}dV+\int\limits_{\partial
V_{t}} \mathbf{P}dS  \label{pyatn}
\end{eqnarray}
where $\mathbf{f}$ is a density of all external mass forces and $\mathbf{P}$
is a surface stress force represented by a stress tensor $P^{ik}$.

The differential form of the relation (\ref{pyatn}) written in components is
often regarded as equation of motion of a continuous medium in Eulerian
coordinates of the fixed reference system:

\begin{equation}
\frac{\partial \rho v^{i}}{\partial t}+\rho v^{i}\,\frac{\partial v^{k}}{%
\partial x^{k}}=\rho f^{i}+\frac{\partial P^{ik}}{\partial x^{k}}
\label{shestn}
\end{equation}%
where $\mathbf{v}$=$\left\{ v^{i}\right\} $ and
$\mathbf{r}=\left\{ x^{i}\right\} $. Takin into account that the
modified continuity equation (\ref{trinadzat}) can be rewritten in
components as:

\begin{equation}
\frac{\partial \rho }{\partial t}+\rho \,\frac{\partial v^{k}}{\partial x^{k}%
}=0  \label{semn}
\end{equation}%
the equation of motion (\ref{shestn}) takes a more simple form:

\begin{equation}
\rho \frac{\partial v^{i}}{\partial t}=f^{i}\rho +\frac{\partial P^{ik}}{%
\partial x^{k}}  \label{vosemn}
\end{equation}

Importantly to emphasize that no changes were assumed for the analytical
representation of forces in the right-hand side of (\ref{pyatn}). The
modification is concerned only the left-hand side of (\ref{pyatn}) which
refers to the time variation of momentum. Thus, in the traditional approach
the partial time derivative $\frac{\partial v^{i}}{\partial t}$ is replaced
by Euler's derivative $\frac{Dv^{i}}{Dt}$.

Nevertheless, we shall limit our consideration here only by
examples of mass and linear momentum time variations since they
suffice to show the way how the system of fundamental differential
equation of continuum mechanics and other field theories could be
modified. To conclude the Section, we would like to stress another
important feature of the given approach: reconsidered fundamental
differential equations of continuum mechanics can be derived
without any change in their original integral formulation.


\section{Conclusions}

Rational examination of mathematical foundations of continuum
mechanics as being of central importance and wide appeal in
physical field theories, shows serious defect at the very basic
level. To be more specific, we claim to have found that the
traditional demonstration of the important kinematics theorem of
continuum mechanics known as Convection (or Transport) Theorem
does not take into account a special constraint on integrand
functions given in Lagrangian representation and, as a
consequence, it is not already based on a mathematically rigorous
approach.

Any modification of the conventional procedure would imply
undeniable changes in the set of basic differential equations of
continuum mechanics as well as some other autonomous branches of
physical science such as electromagnetism, magnetohydrodynamics,
plasma physics etc. Moreover, in this work we show that these
modifications would not be accompanied by any change in
corresponding integral formulations, leaving them untouched. The
latter fact is important from the practical meaning, since the
prevailing amount of experimental data in physical field theories
was basically classified in form of integral laws. Thus, the major
point that emerges from the above considerations is that the
traditional transition from original integral formulations of
physical laws of classical field theories to their differential
form may come in conflict with the mathematical rigor.


$$$$ {\large {\bf REFERENCES}}

\begin{enumerate}
\item{} M. Kline, \textit{Mathematical Thought from
Ancient to Modern Times}, Vol. 2 (Oxford University Press, New
York, 1972)

\item{} R. Smirnov-Rueda, \textit{Found. Phys.}, \textbf{35}(10)
(2005)

\item{} B. Dubrovin, S. Novikov and A. Fomenko, \textit{Modern
Geometry}, Vol. 1 (Ed. Mir, Moscow, 1982)

\item{} R.E. Meyer, \textit{Introduction to Mathematical Fluid
Dynamics} (Wiley, 1972)

\end{enumerate}

\end{document}